\def\MM{M} 
\def\MMu{M} 
\def\mm{m} 
\def\mme{m} 
\global\long\def\order#1{\mathcal{O}\left(#1\right)}

\documentclass[letterpaper,english,prd,preprintnumbers,nofootinbib,showpacs,shortbibliography,notitlepage]{revtex4-1} 
\usepackage[utf8]{inputenc}
\usepackage{amsmath}
\usepackage{nicefrac}
\usepackage{graphicx}

\begin{document}
\preprint{Alberta Thy 1-21, P3H-21-026, TTK-21-13}
\title{Three-loop corrections to the muon and heavy quark decay rates}
\author{Micha{\l} Czakon}
\affiliation{Institut für Theoretische Teilchenphysik und Kosmologie,
  \\
RWTH Aachen University, D-52056 Aachen, Germany}
\author{Andrzej Czarnecki }
\affiliation{Department of Physics, University of Alberta, Edmonton,  Alberta, Canada T6G 2E1}
\author{Matthew Dowling}
\affiliation{Department of Physics, University of Alberta, Edmonton,  Alberta, Canada T6G 2E1}

\begin{abstract}
Recently, $\order{\alpha^3}$ corrections to the muon decay 
rate and $\order{\alpha_s^3}$ to  heavy quark decays have been
determined by Fael, Sch\"onwald, and Steinhauser. This is the first
such perturbative improvement of these important quantities in more
than two decades. We reveal and explain a symmetry pattern in these
new corrections, and confirm the most technically difficult parts of
their evaluation.
\end{abstract}
\maketitle

\section{Introduction}
Muon decay provides one of the pillars of the Standard Model, the
Fermi constant $G_F$ \cite{Marciano:1999ih}. Its determination
relies on precise measurements of the muon lifetime
\cite{Webber:2010zf,Tishchenko:2012ie} and mass \cite{Nishimura:2020jjm},
and on theoretical evaluation of radiative corrections. These
corrections arise primarily from quantum electrodynamic (QED)
interactions involving the muon and the daughter electron, and are
expressed as a power series in the fine structure constant $\alpha
\simeq 1/137$. 

Very recently, a new term in this series has been calculated
\cite{Fael:2020tow} using an expansion in the difference of the muon
and electron masses.  
Here we confirm the most demanding parts of the expansion
published in  \cite{Fael:2020tow}. We also explain a pattern governing
the first two terms of the expansion. 

Determination of the coefficients of increasing powers of $\alpha$
requires monumental efforts. A new result appears very rarely,
approximately once every one or two generations of theorists. Each new result is witness to a new
technology available in perturbative quantum field theory. First order
corrections, calculated in 1955 \cite{Behrends:1955mb}, were in fact
the first loop effects calculated for a decay process. They served as
a model for subsequent studies of quantum chromodynamics (QCD)
processes in heavy quark decays
\cite{Jezabek:1988iv,Czarnecki:1994pu}.

 It took more than 40 years before
the second-order coefficient became known
\cite{vanRitbergen:1999fi,vanRitbergen:1998yd}. That progress was made
possible by the technique of recurrence relations based on 
integration-by-parts \cite{Chetyrkin:1981qh,Tkachov:1981wb} and on
symbolic manipulations with computers \cite{vanRitbergen:1996ng}.

Now, another two decades later, the coefficient of $\alpha^3$ has been
determined \cite{Fael:2020tow}. Several theoretical  developments
underly this advancement. Laporta algorithm
\cite{Laporta:2001dd} allows to express multi-loop integrals
in terms of a relatively small number of master integrals. This
algorithm can be implemented in powerful symbolic algebra software
capable of parallelization
\cite{Ruijl:2017dtg,Smirnov:2019qkx}. Finally, despite the electron being about
207 times lighter that the muon, the calculation is done as an
expansion around the situation where the electron and the muon masses,
$m_e$ and $m_\mu$,
are equal \cite{Archambault:2004zs,Czarnecki:1996gu,Dowling:2008mc}.

Historically, radiative corrections to the muon decay were calculated
before corrections to heavy quark decays, because QCD was developed
only later. Computational technology is the same in both QED and
perturbative QCD,
and once the QED corrections are known, it is relatively easy to
evaluate additional diagrams involving multi-gluon
vertices. Ref.~\cite{Fael:2020tow} presents corrections both for the
muon and for the $b$-quark decays. Wherever our discussion below is
relevant for both types of processes, we refer to the decaying
particle's mass as $M$, and denote the mass of the produced charged
particle by $m$. 
We use $\delta = 1-m/M$ to denote the expansion parameter: the relative
difference of masses.

In Section \ref{sec:exp} we present our calculation of five terms of
the expansion in $\delta$ for a subset of corrections, including the most
demanding part with all quanta (photons or gluons) coupling to
incoming or outgoing
fermions, rather than, for example, to virtual loops.
In Section \ref{sec:ff}  we demonstrate how the first two terms in
the expansion in $\delta$ can be found from a form
factor, first determined in 2004 in Ref.~\cite{Archambault:2004zs}. 
Section \ref{sec:con} contains our conclusions.

\section{Expansion in the mass difference \label{sec:exp}}

As we shall demonstrate in Section \ref{sec:ff}, the first two terms
of the expansion of the decay rate in the mass difference may be
obtained without a new calculation. Unfortunately, the remainder of
the expansion requires a challenging calculation.  For this
publication, we have evaluated the first five terms of the expansion,
$\delta^{5,\dots,9}$ of the three most difficult contributions: those
proportional to $C_F^3$, $C_F^2 n_b$,  and  $C_F n_b^2$ where
$n_b$ labels loops containing a fermion of mass $M$. (We use the
standard notation for the SU($N$) factors: $C_F=(N^2-1)/(2N)$, $C_A=N$,
$T_F=1/2$ with $N=3$ for QCD and $C_F=1$, $C_A=0$, $T_F=1$ for QED
\cite{Muta:2010xua}.) Our method, originally developed for corrections
of $\order{\alpha^2}$ or  $\order{\alpha_s^2}$ in Ref.~\cite{Dowling:2008mc}, is the same as
the one employed in Ref.~\cite{Fael:2020tow} for
$\order{\alpha_s^3}$ ($\alpha_s$
denotes the strong coupling constant). However, we used a different software
implementation.

In this section we briefly describe our method using the example of
the muon decay.  It was found in Ref.~\cite{Dowling:2008mc} that an
expansion in terms of $\delta$ is useful for computing high order
corrections. In particular, the calculation is much less involved
than an expansion around $m/M=0$, and converges very well all the way to
the physical value of ${m_c/m_b}$ or ${m_e}/{m_\mu}$, even
though $m_e/m_\mu=0.005$ is very far from the equal mass limit.

Our crucial tool is the optical theorem.
It allows us to write contributions to the decay rate of the
$b$-quark or muon as self energy diagrams. To generate the diagrams,
we use \textsc{DiaGen/IdSolver} \cite{diagen} modified to work with
the types of propagators and expansions that appear in the present calculation.
This code also produces the required integration by parts (IBP) identities
and carries out the reduction to master integrals. The reduction of
the large number of necessary integrals  required
the use of the WestGrid cluster which is part of the high performance
computing infrastructure Compute Canada.

\begin{figure}[htb]
  \begin{center}
    \includegraphics[width=.3\textwidth]{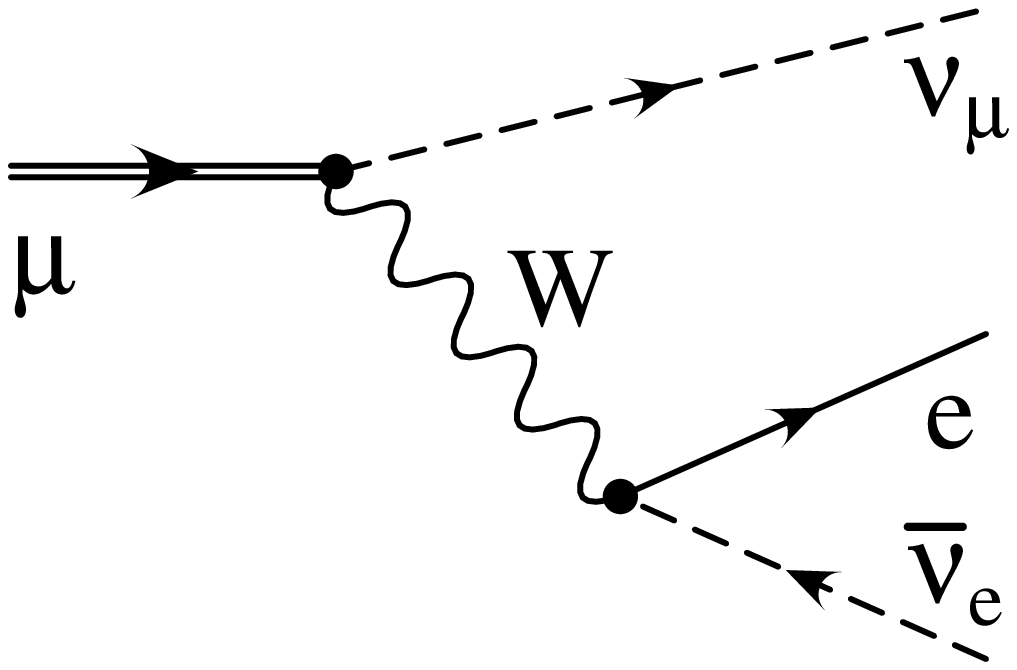}
    \hspace*{20mm}
    \includegraphics[width=.35\textwidth]{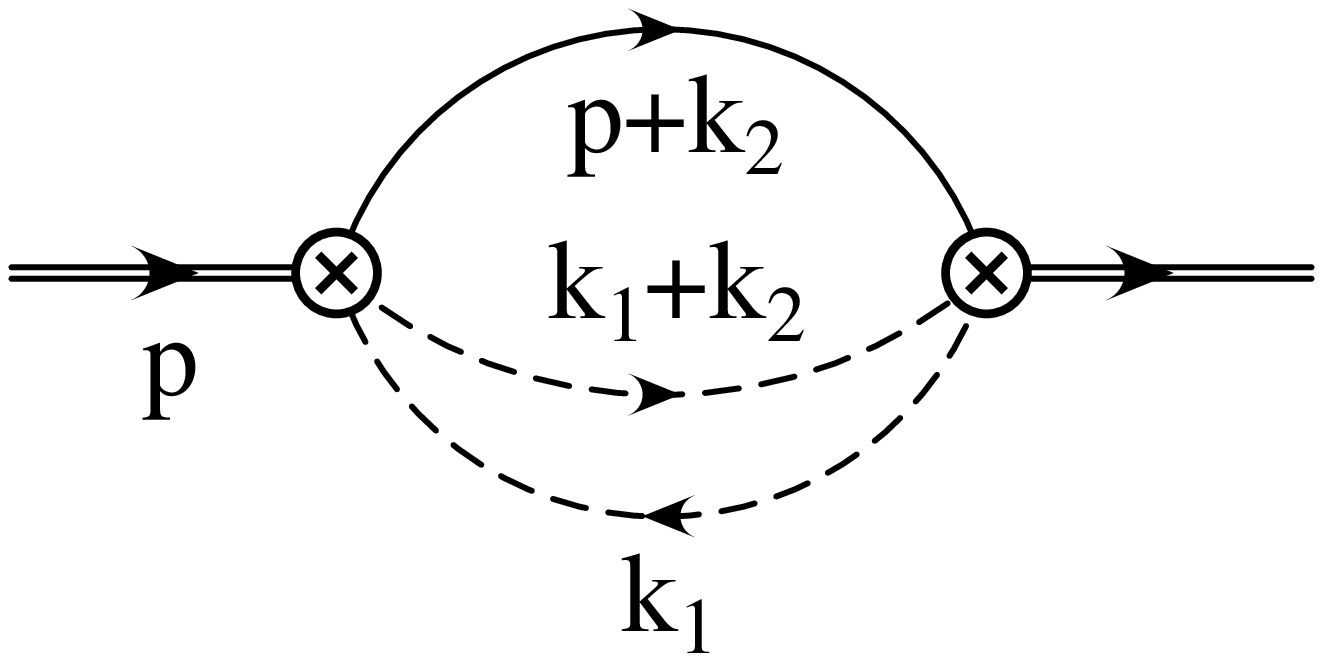}
\\
     \hspace*{-10mm}
 (a) 
    \hspace*{77mm}
 (b) 
    \caption{\label{fig:tree} The tree level contribution to muon
      decay. Panel (a) shows the full amplitude with the $W$ boson. 
Panel (b) shows the muon self-energy diagram whose imaginary part we
calculate to obtain the squared amplitude. Circled crosses indicate effective
interactions obtained by neglecting the momentum dependence in $W$
propagators (Fermi interaction).}
  \end{center}
\end{figure}

\begin{figure}[htb]
  \begin{center}
    \includegraphics[width=.4\textwidth]{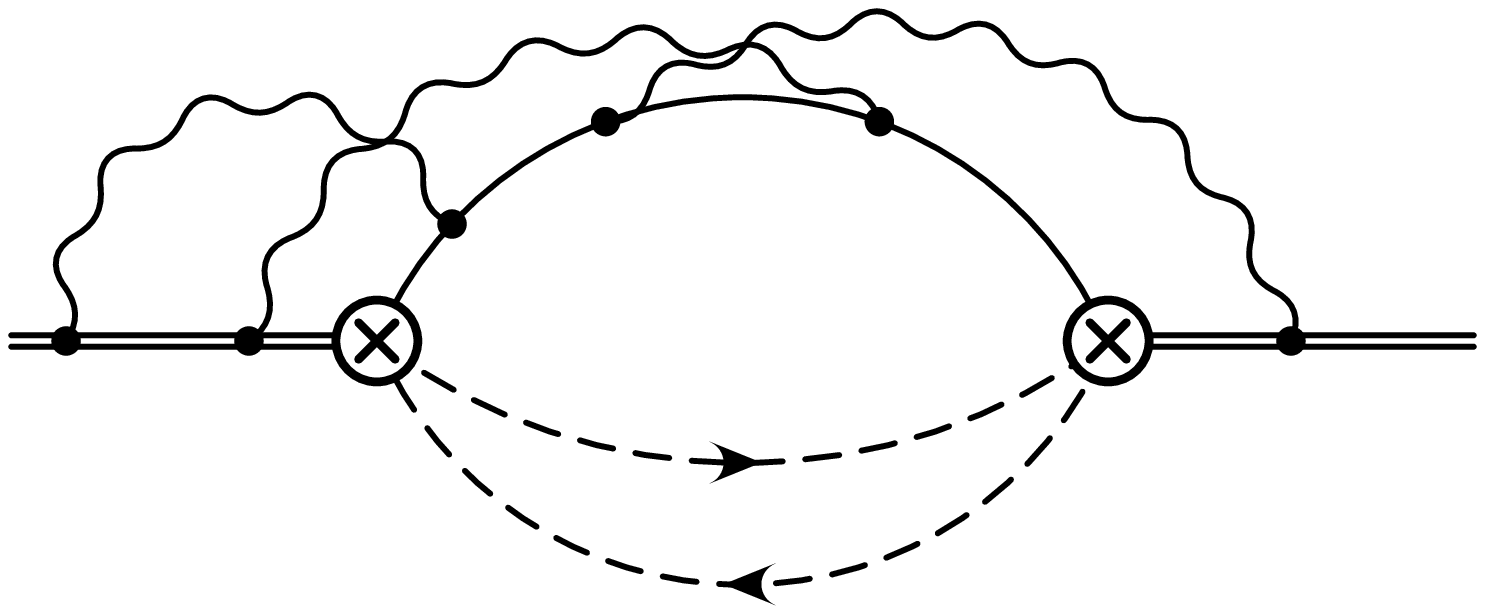}
    \hspace*{3mm}
    \includegraphics[width=.4\textwidth]{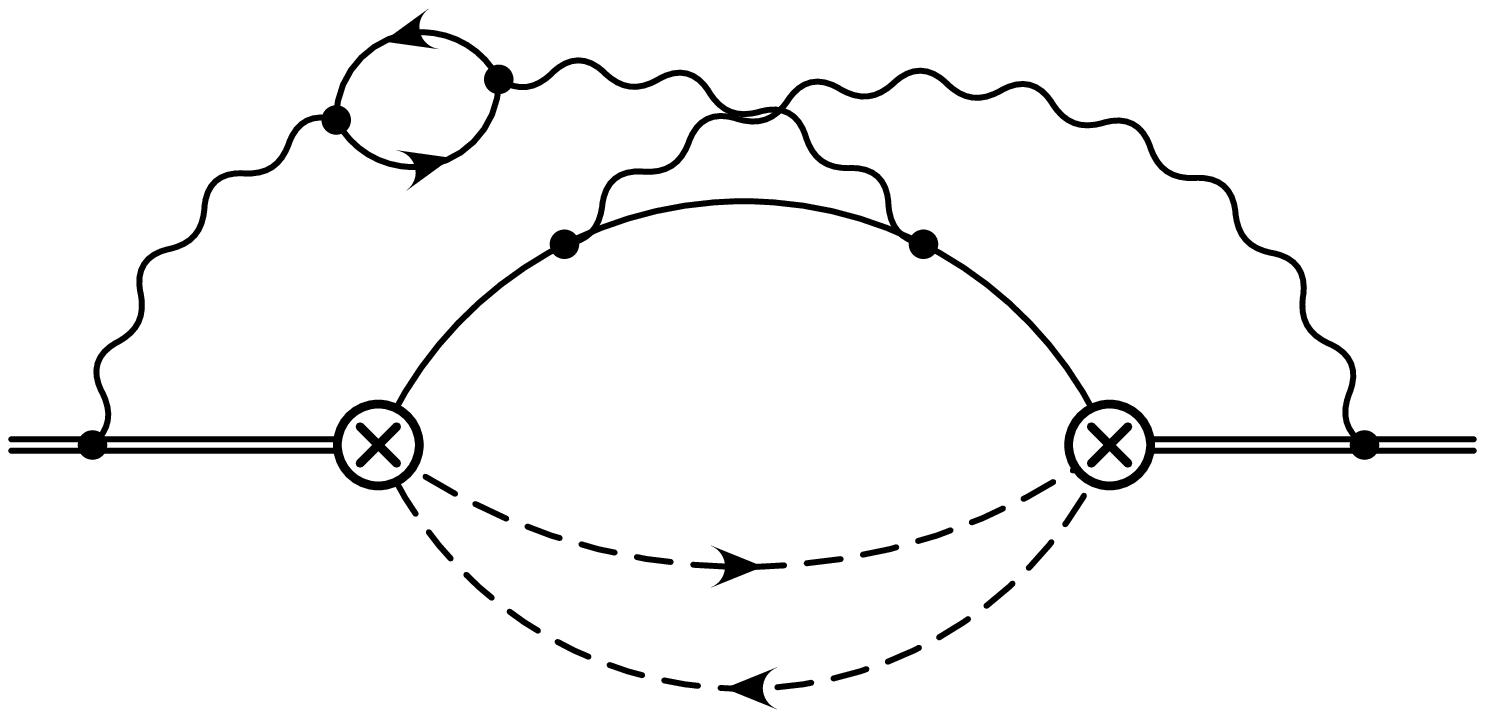}
\\  
 (a)     \hspace*{72mm} (b)
\\
    \includegraphics[width=.4\textwidth]{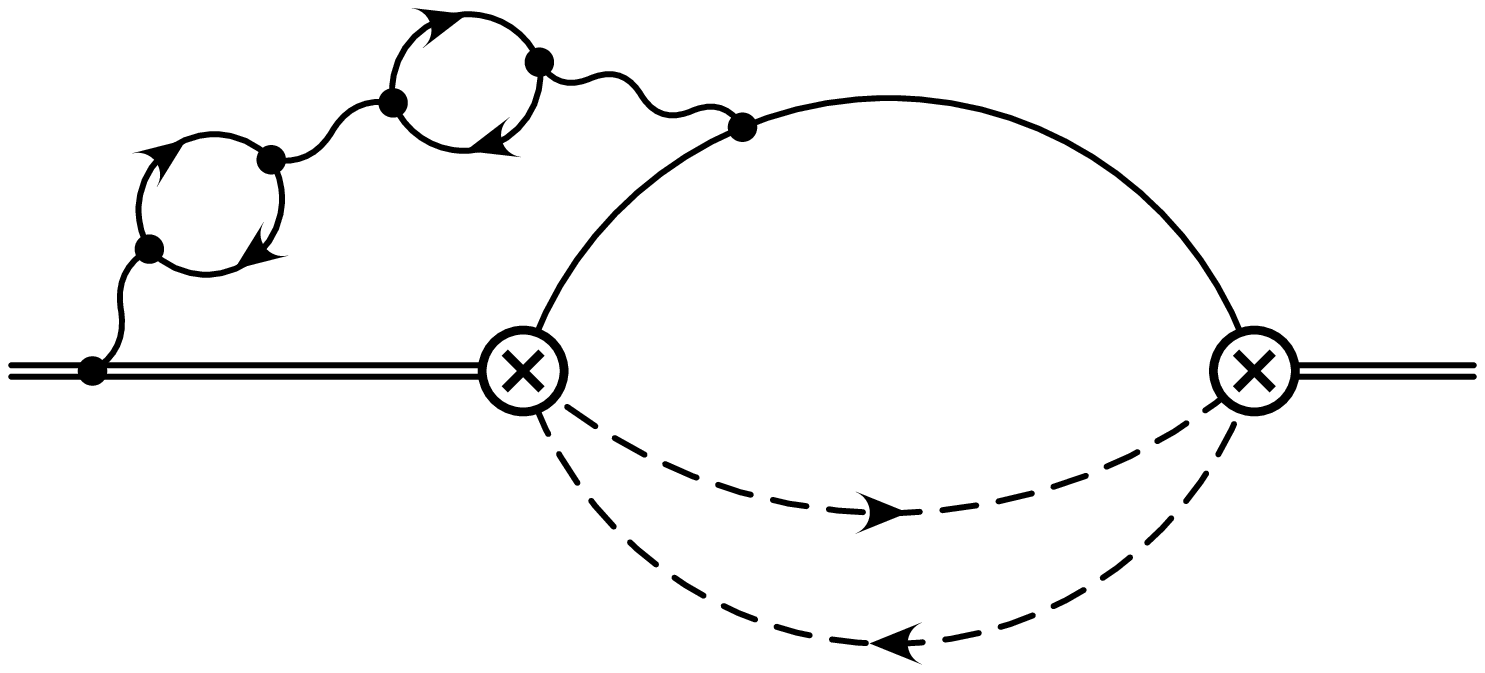}
\\   (c) 
    \caption{\label{fig:RC} Examples of third-order radiative
      corrections to the muon decay. Wavy lines represent photons,
      contributing to both virtual and real radiation. In our
      approach, the imaginary part of such five loop diagrams is calculated. }
  \end{center}
\end{figure}

The decay is induced by an effective Fermi interaction with the $W$
boson contracted to a point, as shown in Fig.~\ref{fig:tree}.
 Corrections due to the $W$ propagator including electroweak effects
are known up to two loops \cite{Awramik:2003ee}. Here we are
interested in three loop QED corrections in the limit of a heavy
$W$. They are described by five loop self energy diagrams (the extra
two loops are responsible for the tree-level decay in
Fig.~\ref{fig:tree}(b)). We show examples of such diagrams in Fig.~\ref{fig:RC}.
The integral over the neutrino loop is carried out first by
integrating over $k_{1}$ shown in Fig.~\ref{fig:tree}(b). This can be done at
any QED order because the neutrinos do not interact with each other in
the approximation considered here.  The number of loops is thus
reduced by one, at the price of introducing a propagator with
non-integer power. This complicates the IBP reduction.

For the remaining four loop integrals, we use an asymptotic expansion
in $\delta$, which we now describe.

\subsection{Asymptotic Expansion}
\label{asExp}
General properties of the asymptotic expansion of the loop integrals
can be explained with the help of Fig.~\ref{fig:tree}. Since we have
already integrated over the neutrino loop, we only need to
consider $k_{2}$ when carrying out the expansion. The
following integral remains
\begin{equation}
\int\frac{d^d k_2 }{(2\pi)^d }\frac{1}{(k_2 ^2 )^\epsilon 
[(p+k_2 )^2 -\mm^2 ]} 
=  \int\frac{d^d k_2 }{(2\pi)^d }\frac{1}{(k_2 ^2 )^\epsilon 
[k_2^2 + 2 p\cdot k_2 +\MM^2\delta(2-\delta) ]} 
\label{toExp}
\end{equation}
 where we use dimensional regularization with $d=4-2\epsilon$. Without
 QED corrections, there  are two regions: soft with components $k_2 $
 of the order of $\MM-\mm$, thus 
 much smaller than $\MM$;  and hard with $k_2 \sim  \MM$. In the hard region,
 the term $\MM^2\delta(2-\delta)$ can be treated as small. After
 expanding in it, the integral takes the form
\begin{equation}
\int\frac{d^d k_2 }{(2\pi)^d }\frac{1}{(k_2 ^2 )^{\epsilon}\left[k_2
    ^2 +2p\cdot k_2 \right]}.
\end{equation}
 This is an on-shell self-energy integral with no imaginary part and
 so it does not contribute
to the decay. This is the crucial observation which holds in any order
in QED: if the momentum
flowing through the neutrinos is large, the resulting integral
has no imaginary part. This reduces the
number of regions that must be considered by half. At $\order{\alpha^3}$
this means that we only need to consider eight regions instead of the
general 16. Moreover, among the non-contributing regions is the computationally very
difficult one with all five hard loops.

In the soft region, $k_2 \sim \delta\cdot \MM$ and 
we simplify the integral in Eq.~(\ref{toExp}) by noticing that some
terms are suppressed by an extra power of $\delta$,
\begin{equation}
\int\frac{d^d k_2 }{(2\pi)^d }\frac{1}{(k_2 ^2 )^{\epsilon}[2p\cdot k_2 +2\MM ^2 \delta+\underbrace{k_2 ^2 -\MM ^2 \delta^2 }_\text{small}]}.
\end{equation}
Thus we encounter integrals  with so called eikonal propagators \cite{Bagan:1992ty,Czarnecki:1996nr}
whose general form is
\begin{equation}
\int\frac{d^d k}{(2\pi)^d }\frac{1}{(k^2 )^{n_{1}}[2p\cdot k]^{n_2
  }\left[2p\cdot k+\Delta\right]^{n_{3}}},
\label{eq:eik-one}
\end{equation}
 where $\Delta$ does not depend on the integration variable $k$.
We have included another eikonal propagator
that appears in  higher-order loop diagrams. As it turns out,
in all cases $\Delta$ takes the form of a constant or another eikonal
propagator. This makes it possible to apply this integral iteratively
when computing the required master integrals.

\subsection{Master Integrals}

Expansion in relevant momentum regions of all diagrams leads to  a large number of eikonal and on-shell
integrals. We encounter at most three-loop on-shell integrals
because the four-loop on-shell integrals 
contribute only to the real part and not to the imaginary part that we
need.  

Fortunately the three-loop on-shell master integrals have been
computed in Ref.~\cite{Lee:2010ik}  to the required order in
$\epsilon$ for the present calculation. We note that these integrals
were originally computed in Ref.~\cite{Laporta:1996mq} in the context
of the fermion anomalous magnetic moment, $(g-2)$, but it was not
until \cite{Lee:2010ik} that enough terms in the $\epsilon$ expansion
where known to make the present calculation possible.

The soft integrals appear up to four loops. After reducing
to master integrals, there are one one-loop, one two-loop, three three-loop
and 13 four-loop master integrals that must be computed. The general
one-loop eikonal integral in Eq.~\eqref{eq:eik-one} is computed
in \cite{Bagan:1992ty}. The result has the property
\begin{equation}
\int\frac{d^d k}{(2\pi)^d }\frac{1}{(k^2 )^{n_{1}}[2p\cdot k]^{n_2
  }[2p\cdot k+\Delta]^{n_{3}}}
\propto \Delta^{D-2n_{1}-n_2 -n_{3}}.
\end{equation}
As discussed in Section \ref{asExp}, the form of $\Delta$ in this
result makes it possible to iteratively apply the one loop computation
to all two and three-loop and almost all four-loop eikonal type master
integrals that are required.  The remaining four-loop eikonal
integrals can be computed analytically using a single Mellin-Barnes
parameter. In all cases, the Mellin-Barnes integral can be evaluated
by summing over the poles of the Gamma functions.  Thus all eikonal
integrals can be written explicitly in terms of Gamma and
hypergeometric functions which are then expanded in $\epsilon$ to the
needed order. For the expansion of hypergeometric functions, we use
the Mathematica package \textsc{HypExp} \cite{Huber:2007dx}. Detailed
examples of evaluation of eikonal integrals can be found in
Ref.~\cite{Fael:2020njb,Dowling:2012phd}. 

The final ingredient are the three-loop renormalization constants.
The mass dependent constants are known in an expansion around the
limit ${\mm/\MM}=0$ \cite{Bekavac:2007tk}.  We require the opposite
expansion and thus had to compute the wave-function and mass
renormalization terms. The calculation was done in an identical way to
\cite{Bekavac:2007tk} except, of course, expanding around $\mm/\MM=1$.

Interestingly, authors of \cite{Bekavac:2007tk} were able to present
their solution only in terms of one- and two-dimensional integrals
which must be computed numerically. In the limit $\mm/\MM=1$ however,
we obtain fully analytic results. [After this calculation
was completed, we learned about Ref.~\cite{Fael:2020bgs} where  analytic results in the
limits $\mm/\MM=0,1,\infty$ are presented.]

Using the approach described in this section, we have been able to
confirm terms in the first five orders, $\delta^5, \dots, \delta^9$, of 
Ref.~\cite{Fael:2020tow}, in the most technically difficult groups:
the ones that contain three photons, all coupling to the main
muon-electron line, one example of which is shown in
Fig.~\ref{fig:RC}(a);  and diagrams containing one or two virtual loops with
mass $M$, as shown in Fig.~\ref{fig:RC}(b,c).

 In principle, an extension
of our calculation to diagrams containing virtual $m$ loops and
massless loops, and three-gluon vertices is possible. Instead, in the
following Section \ref{sec:ff}, we show how one can reproduce the
first two terms of all types from already known form factors.

\section{Information from the zero-recoil form factor \label{sec:ff}}
\subsection{Hidden symmetry under $M\leftrightarrow m$}
The starting point of the expansion in the difference of masses is the
equal mass limit. There, the three loop correction has been known for
a long time \cite{Archambault:2004zs}. That result can be used to
obtain the first two terms in the mass difference expansion, as
explained in Ref.~\cite{Dowling:2008mc}.

Without radiative corrections, the effect of the electron mass on the
muon decay rate is known exactly,
\begin{equation}
  \label{eq:1}
  \frac{\Gamma(\mu\to e\nu\bar \nu)}{\Gamma_{0}} 
=1-8\rho^2 +8\rho^{6}-\rho^{8}-24\rho^{4}\ln\rho,
\end{equation}
where $\Gamma_{0}=\frac{G_{F}^2 \MMu ^{5}}{192\pi^{3}}$ is the
decay rate in the limit of a massless electron and
$\rho=\mme /\MMu $ is the ratio of electron and muon masses. Near
the limit of equal electron and muon masses a useful parameter is
$\delta=(\MMu  - \mme )/\MMu  = 1 - \rho$, introduced in Section \ref{sec:exp}. The first two terms in
the expansion of the tree-level decay rate around $\delta = 0$ are
\begin{equation}
  \label{eq:2}
   \frac{\Gamma(\mu\to e\nu\bar \nu)}{\Gamma_{0}}  = \delta^5 - {3
     \over 2}\delta^6 + \order{\delta^7}.
\end{equation}
We see that the phase space suppresses the decay rate by five powers
of $\delta$. 

The following discussion applies equally to the muon decay  and to the
semileptonic $b\to c$  and   $b\to u$ quark  decays. Quark decays
require a more extensive calculation because
QCD corrections have a richer gauge group structure. We therefore use the
$b\to c$ decay as an example, use the
language of QCD, and denote $\delta  =(m_{b} -
m_{c})/m_b$. Corrections to the muon decay can be deduced from a
subset of results obtained for quarks.

We want to argue that when radiative corrections are included,
the first two orders in $\delta$, displayed in Eq.~\eqref{eq:2}, are affected only by virtual
corrections and not by the real radiation. Real radiation emission is
suppressed by two powers of the daughter fermion velocity in the rest frame
of the decaying fermion. That velocity, $v=\sqrt{1 - {m_c^2 \over
    m_b^2}}$, is at most about $\delta$. So, the real radiation
affects the decay rate only in the third order in the expansion,
$\delta^7$.

It has also been demonstrated that the form factors arising from
virtual corrections do not have a linear term in the $\delta$
expansion \cite{Dowling:2008mc}, as long as the coupling constant
$\alpha_s$ is renormalized in the symmetric point $\mu^2 = m_c
m_b$. This is because of the symmetry of the virtual gluon diagrams
under the interchange $m_b \leftrightarrow m_c$. We have therefore used
Eq.~\eqref{eq:2} to reproduce the first two terms of the expansion in
\cite{Fael:2020tow}. 

The available information about the form factors in
\cite{Archambault:2004zs} is limited to the axial form
factor. Fortunately, in the limit of equal masses, the vector form
factor does not receive radiative corrections. Corrections are
suppressed by the square of momentum transfer $q^2 = (p_b - p_c)^2$
which contributes again only in the third order, $\delta^7$. 

Since the results of \cite{Archambault:2004zs}  assume
the exact equal mass limit, they do not differentiate between the two
heavy fermions. In Ref.~\cite{Fael:2020tow} effects of $b$ and $c$
loops are labeled with factors $n_b$ and $n_c$, to indicate that they
are proportional to the number of those quarks in virtual loops (in
reality of course there is $n_b=n_c=1$). We thus replace $n_b$ and $n_c$ in the formulas of
\cite{Fael:2020tow} by $N_H/2$, so that the number of heavy fermions
$N_H$ in \cite{Archambault:2004zs} corresponds to $n_b+n_c$. 

Finally,
when we apply Eq.~\eqref{eq:2}, we obtain corrections in terms of the
coupling constant normalized at the geometric mean of the decaying and
daughter fermion. On the other hand, results of
Ref.~\cite{Fael:2020tow} use the mass of the decaying fermion as the
renormalization scale. To remedy this, we run the coupling constant in
the formulas resulting from Eq.~\eqref{eq:2} using
\cite{Kniehl:2002br},
\begin{align}
  \label{eq:3}
 \frac{\alpha_s(m_bm_c)}{\pi} =&
\frac{\alpha_s(m_b^2)}{\pi}\left[1+\frac{\alpha_s(m_b^2)}{\pi}\beta_0\delta
+\left(\frac{\alpha_s}{\pi}\right)^2\delta
                               \left( \beta_0^2\delta+\beta_1\right)\right],
\\
\beta_0=&{1\over4}\left({11\over3}C_A-{4\over3}T_Fn_f\right),
\\
\beta_1=&{1\over16}\left({34\over3}C_A^2-{20\over3}C_AT_Fn_f-4C_FT_Fn_f
\right),
\end{align}
where $n_f = n_l+N_H$. In the above formulas, $\delta$ originates
from the logarithm $\ln {m_b\over m_c} = - \ln (1-\delta) = \delta +\order{\delta^2}$.

After this adjustment of the coupling constant in the axial form
factor, we reproduce all coefficients of $\delta^5$ and $\delta^6$ in
Ref.~\cite{Fael:2020tow}.

\subsection{Example of the pattern in terms $\delta^5$ and $\delta^6$}
As an example of the procedure outlined above, we consider the first
two terms in the color part proportional to $C_F C_A^2$. In the Supplementary
Material of Ref.~\cite{Fael:2020tow}  these terms read $C_F C_A^2
\left(d_5 \delta^5  + d_6 \delta^6 \right)$, with
\begin{align}
d_5 & =
  -{16241 \over 270} - {512 \over 15}a_4 - {64 \over 45}\ln^4 2 - {1423 \over 90}\pi^2 + 
   {1112 \over 45}\pi^2 \ln 2- {40 \over 9}\pi^2 \ln^2 2+ {194 \over 675}\pi^4 + {86 \over 3}\zeta_3 + 
   {22 \over 15}\pi^2\zeta_3 - 32 \zeta_5,\\
d_6 & ={1745 \over 36} + {256 \over 5}a_4 + {32 \over 15}\ln^4 2 + {1247 \over 60}\pi^2 - 
   {156 \over 5}\pi^2 \ln 2+ {20 \over 3}\pi^2 \ln^2 2- {97 \over 225}\pi^4 - {259 \over 5}\zeta_3 - 
   {11 \over 5}\pi^2\zeta_3 + 48 \zeta_5,
\end{align} 
where $a_4 = \text{Li}_4(1/2)\simeq 0.517$ \cite{lewin} and $\zeta_n$
is the Riemann zeta function \cite{NISTHandbook}.
Following the discussion around Eq.~\eqref{eq:2}, the bulk of 
these terms should have the form $d_5\delta^5(1-3\delta/2)$. Consider the
deviation from that pattern,  $D=d_6 + 3d_5/2$,
\begin{equation}
D = 
 - {1879 \over 45} - {44 \over 5} \zeta_3 - {44 \over 15} \pi^2 + {88
   \over 15}  \pi^2 \ln 2,\label{dev}
\end{equation}
a much simpler expression than either $d_5$ or $d_6$, but not zero.

We now show how this difference is removed by changing the
renormalization scale. To this end, consider one- and two-loop form
factor
\cite{Shifman:1987rj,Paschalis:1982zq,Close:1984ad,Czarnecki:1996gu},
as summarized in \cite{Archambault:2004zs},
\begin{align}
\eta_A(\delta \to 0) &= 1 + \eta_A^{(1)}{\alpha_s(m_bm_c)\over \pi}
+ \eta_A^{(2)}\left(\alpha_s( m_bm_c)\over \pi\right)^2 +\order{\alpha_s^3},\\
\eta_A^{(1)} (\delta \to 0) &=  -{C_F \over 2\pi} \\
\eta_A^{(2)} (\delta \to 0) &= C_F C_A  \left( - {143 \over 144} - {1
    \over 4} \zeta_3 + {1 \over 6} \pi^2 \ln 2 - {1 \over 12}
  \pi^2\right) + \dots,
\end{align}
where we have neglected other color structures in $\eta_A^{(2)} $.
Change of  the scale of $\alpha_s$ in the square of the form factor produces an
additional contribution $\alpha_s^3C_FC_A^2$,
\begin{align}
{48\over 5} \eta_A^2 & \to {96 \over 5} 
\left[ \eta_A^{(1)}  \beta_1 +2 \eta_A^{(2)}\beta_0\right]\\
& \to  {96 \over 5} 
\left[ -{1\over 2}\cdot   {17 \over 24} 
+2{11\over 12} \left( - {143 \over 144} - {1 \over 4} \zeta_3 + {1 \over 6} \pi^2
   \ln 2 - {1 \over 12} \pi^2\right)  \right] = D, \label{agree}
\end{align}
where the factor $48/5$ is related to the normalization of the results
in \cite{Fael:2020tow}. The result of changing the scale from the
symmetric point $m_bm_c$ to $m_b^2$, given by  Eq.~\eqref{agree}, provides
precisely the deviation from the pattern $d_5\delta^5(1-3\delta/2)$ we found in Eq.~\eqref{dev}.

\section{Conclusions \label{sec:con} }
The third order correction to the muon and the heavy quark decay rates
found in Ref.~\cite{Fael:2020tow} is an important milestone in
perturbative quantum field theory. Here we have confirmed the first five terms
of the three technically most difficult structures involving no light vacuum
polarization loops and no three-gluon couplings. 
We have also revealed a pattern in the first two terms of the
expansion of all flavor and color structures, hidden when the symmetry
between the initial and the final fermion is broken by
renormalization. This pattern becomes visible when a symmetric scale
$Mm$ is used to renormalize the coupling.

We close by congratulating the authors of Ref.~\cite{Fael:2020tow} on
completing their heroic calculation.

\begin{acknowledgments}
A.~C.~thanks Matthias Steinhauser for helpful discussions. 
The work of M.~C.~was supported by the Deutsche Forschungsgemeinschaft
under grant 396021762 - TRR 257. 
The work of A.~C.~and M.~D.~was supported
by the Natural Sciences and Engineering Research Council of
Canada. This research was enabled in part by support provided by
WestGrid (www.westgrid.ca) and Compute Canada (www.computecanada.ca).
\end{acknowledgments}


\end{document}